# Interchromatidal central ridge and transversal symmetry in early metaphasic human chromosome one

Orlando Argüello-Miranda[1,2] and Giovanni Sáenz-Arce[2,3*]

1Escuela de Ciencias Biológicas, Universidad Nacional de Costa Rica.
2Laboratorio de Nanotecnología, Centro Nacional de Alta Tecnología, Costa Rica.
3Laboratorio de Materiales, Departamento de Física, Universidad Nacional de Costa Rica

**Abstract**

The topographic structure of Giemsa banded (G-banded) early metaphase human chromosomes adsorbed on glass was analyzed by atomic force microscope using amplitude modulation mode [AM-AFM]. Longitudinal height measurements for early metaphasic human chromosomes showed a central ridge that was further characterized by transversal height measurements. The heterochromatic regions displayed a high level of transversal symmetry, while the euchromatic ones presented several peaks across the transversal height measurements. We suggest that this central ridge and symmetry patterns point out a transitional arrangement of the early metaphase chromosome and support evidence for interchromatidal interactions prior to disjunction.

## INTRODUCTION

Chromosomes are vehicles of genetic information. Their structure and organization have been the focus of investigation before DNA was known to contain inheritance information. Nowadays nanometric resolution techniques like Scanning Probe or Electronic Microscopy are used to image the chromosome configuration; several captions of ultrastructures using these techniques have contributed to establish a less theoretical approach to the real conformation of the chromosomes [Harrison *et al*. 1982; Kenneth *et al*. 1986; Stark *et al*. 1998; García & Pérez, 2002]. According to these studies it is now accepted that the metaphasic chromosomes are super coiled strings of DNA and proteins, which have several hierarchical levels of organization [Cummings, 2003; Grigoryev, 2004].

Since Atomic Force Microscopy (AFM) has been used to image human chromosomes, a more accurate documentation on the three dimensional structures of chromosomes has been developed. The height, width, volume, chromosomal aberrations, translocations, deletions and other features are already reported for some karyotypes and specific chromosomes [Rasch *et al*. 1993; Jiao & Schäffer, 2004; Zerrin *et al*. 2003] furthermore AFM images have been achieved either in native conditions, i.e. in liquid, without previous treatment [Hoshi *et al*. 2006], or with different staining methods, giving parameters to detect even the chromosome's damage caused by analysis techniques [Yangzhe *et al*. 2006; Kimura *et al*. 2004]. The classical chromosomal G-banding pattern is depicted with AFM as ridges and grooves, which can be correlated with the DNA A-T and G-C rich portions respectively [Tamayo & Miles, 2000]. The centromeric region and centromeric banding patterns (C-banding) are also defined in the literature [Fukushi & Ushiki, 2005].

Some polemical structures also have been reported. Most of them can be classified as "artifacts" from preparation methods or debris from the sample, in that regard "ghost strands" and satellite chromosomal ultrastructures can be distinguished [Hoshi & Ushiki, 2001]. However, other structures, like interchromatidal fibers, have been regarded as a normal part of chromosomes in some metaphasic stages, connecting the ridged regions in sister chromatids [Ushiki *et al*. 2002].

Fibers bridging the gap between sister chromatids have been reported since 1976 in TEM images from chromosomes replicas [Bath, 1976]. These can not be discarded as artificial products of trypsin or pre-TEM treatment; however there is not enough evidenced yet either against or in favor. The AFM images taken so far couldn't provide any strong evidence, basically because their interpretation at that scale needs a stronger biochemical background. Furthermore, the effects of contamination due to artifacts during the stain or fixation treatments could become significant at that level [Tamayo *et al*. 1999; Tamayo & Miles, 2000].

The existence of discrete chromosomal structures such as interchromatidal fibers can represent an important step in the DNA organization from prophase heading towards metaphase, and thus, a hint to the full understanding of the process of chromosomal arrangement. However most of the AFM studies either focus on chromosomes with already split sister chromatids, or simply ignore the differences of the degree of disjunction of the sister chromatids. It could be even argued, that most of the observations of interchromatidal structures seem to be not intended as a research objective, but a casual observation while some other feature was being tested. The aim of this study is to describe, using AFM, new early metaphasic chromosomal features which can give hints on the processes undergoing the separation of sister chromatids.

## II. MATERIAL AND METHODS

**Sample collection and culture conditions:**

6.0 ml arm peripherial blood samples from 4 healthy volunteers around 20 years old ( 2 men, 2 women) were extracted through venipuncture with vacuinaters containing sodium heparin as anticoagulant. The samples were immediately processed *in situ* at room temperature as follows: 1 ml of blood was inoculated in sterile plastic 15 ml tubes containing 10 ml of prewarmed (37°C) RPMI 1640 medium supplemented with 20% fetal calf serum (FCS). The tubes were gently mixed, and placed in a rack in an incubation chamber for 48 hours at 37°C ± 0.1°C in a 5% $CO_2$ atmosphere.

**Metaphasic arrest and extraction**

Lymphocytes were arrested at different stages during metaphase adding 1.0 ml of 0.05 μg/mL colcemid solution ( Gibo Karyomax colcemid, lot No. 15 212-012 ) to the medium and incubated at 37°C± 0.1°C for 30 min.

In order to concentrated and wash the cells, the culture was centrifuged during 8 minutes at 1200 rpm. The supernatant was discarded and the Cell pellet was first resuspended with 1.5 ml prewarmed (37°C) KCl 0.075 M and then 5.5 ml more of KCl were added, following 18 minutes incubation at 37°C. The centrifugation step was repeated, the supernatant discarded and the pellet along with some leftover liquid were fixed adding 7 ml of homemade 3:1 methanol:acetic acid solution. The centrifugation step was repeated with methanol:acetic acid solution two times more until the pellet and supernatant looked clear.

Finally, the pellet was resuspended in 2 ml fixative, ready for immediate slide preparation.

**Slide preparation and Giemsa-banding.**

Two drops of resuspended pellet were added per each new, clean slide (dropping height: ≈5 cm), which were then air dried and checked under phase contrast microscope to assess an adequate cell density.

G-banding was performed incubating the metaphasic spreads slides in 0.25% trypsin solution at 25°C for 20 seconds, washed with sterile saline solution and then 10% Giemsa stain (Gibco Karyomax Giemsa stain improved R6 solution) was applied for one minute.

The slides were washed with sterile saline solution to avoid stain excess, air dried, and put into a glass desiccation chamber for no more than 48 hours until AFM analysis were done. The slides were allowed five minutes in ambient conditions before AFM scans were started. Changes of volume caused by hydration of the chromosomes in the imaging conditions was assumed negligible.

**Collection of data**

Several preparations were examined in order to find metaphasic preparations showing the typical pattern of chromosomes in early metaphase, i.e. with the chromatids starting to split. Those metaphasic spreads with evident cytoplasmic debris, extremely either twisted or merged chromosomes were ignored. Thirteen early metaphasic spreads, from different volunteers, presenting similar level of trypsin digestion were analyzed. The chromosome one from each metaphase spread was used for calculations.

Longitudinal and transversal height measurements were taken and analyzed with the Analyze Panel from Igor pro 5.0. For longitudinal height measurements, the chromosome area was divided in three equidistant longitudinal zones, being 2 the central longitudinal area and 2, 3 the outer ones. The transversal height measurements were taken on two euchromatic bands (p36, q42) and two heterochromatic bands (p21, q31). The height differences within the bands were tested (t-test) against a mean value of 0 (no difference) with a standard deviation equal to that of the corresponding band (fig. 3).

**Atomic force microscopy conditions**

The AFM scans were performed by the MFP-3D system with a 90 μm scanner [Asylum Research, Santa Barbara, CA.], implemented with a TS-150

mechanical vibration isolation table (Table stable, Ltd. HWL Scientific Instruments, GmbH, Germany). The laser alignment was set to the maximum SUM values for the used cantilever

( 5.65±0.05) in the sum and deflection meter. The cantilever was rectangular, made of Si, with a length of 240 µm [AC240TS, Asylum Research]. Its nominal spring constant (2N/m) and the resonance frequency (70 kHz) were confirmed by the thermal tune method incorporated in the Igor-pro software, version 5.0. The force constant and natural oscillation frequency were checked before any scanning, since an unexpected change in these features could represent either contamination or significant abrasion of the tip.

All images, with simultaneous height, amplitude and phase contrast detection, were collected in amplitude modulation mode (tapping mode) in air at 25°C and relative humidity close to 50%. The instrumental resolution was set at 512 x 512 lines and 256 x 256 points. Scan speeds were approximately of 1Hz, with set points between 0.8 and 0.9 V and integral gain set to 20. The resulting data was processed with Igor Pro 5.0 software [WaveMetrics, Inc.]. Under these optimized feedback parameters longitudinal and transversal height measurements from chromosomes were taken.

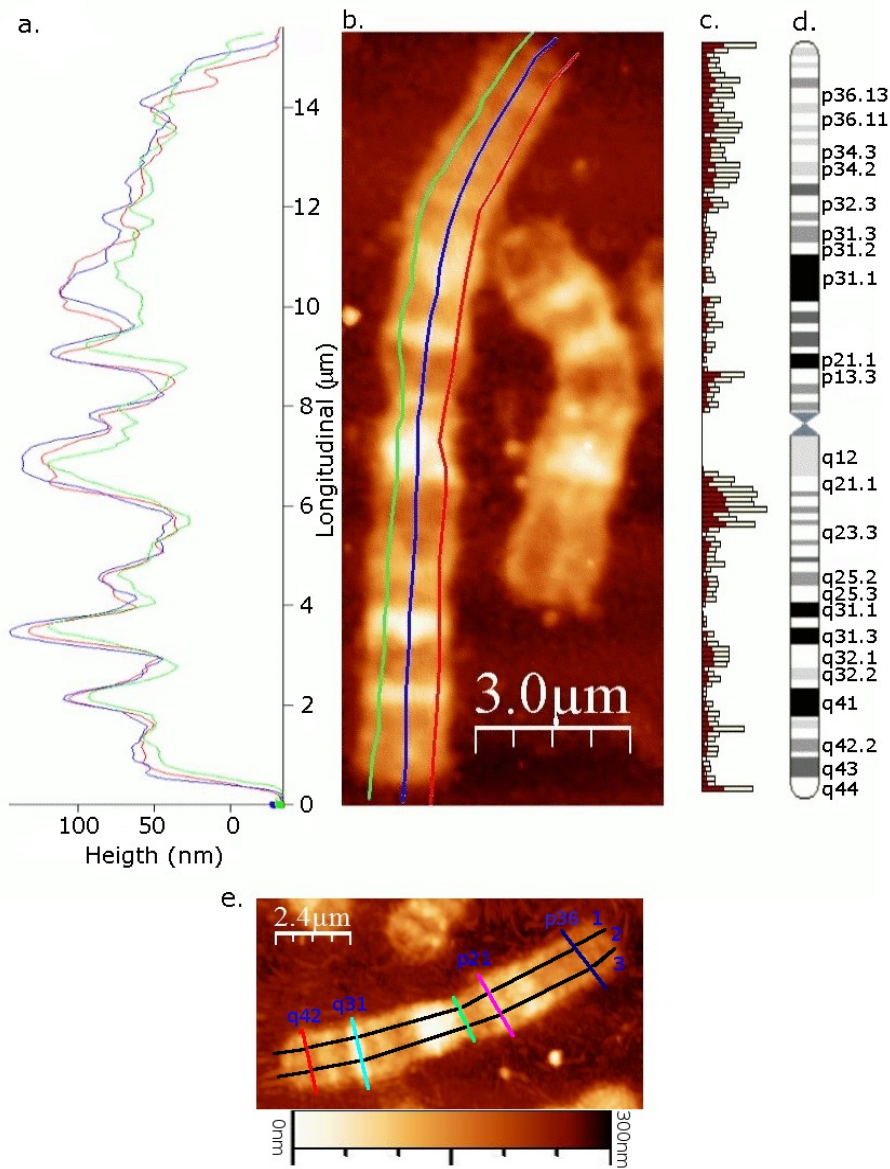

**Figure 1.** Topographic image of G-banded metaphasic human chromosome one a] Longitudinal height measures b] AFM image c] eu- and heteochromatin according to genes density d] *Ensembl's* ideogram used as a control for banding. e] Schematic virtual dissection of a chromosome one showing the central (2) and external (1,3) longitudinal zones and the measured euchromatic( p36, q42) and heterochromatic (p21, q31) bands.

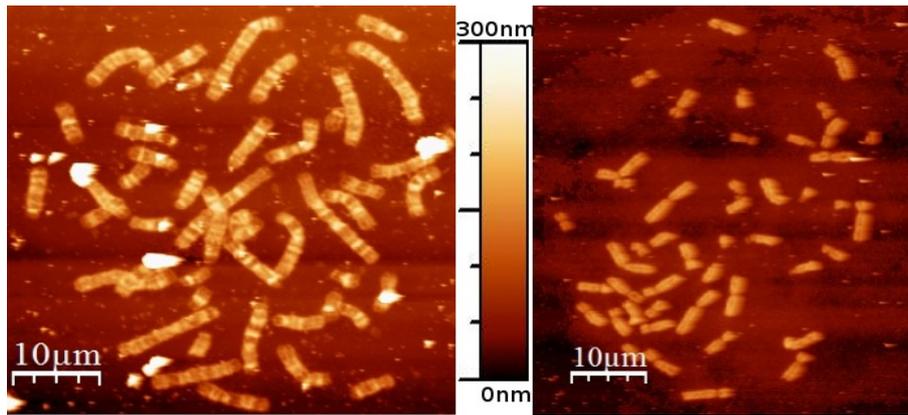

**Figure 2**. Topographic images of two different G-banded metaphasic human chromosomes. [Left] Each chromosome in early metaphase can be distinguished as having two recently built sister chromatids, however both of them are in tight contact previous to disjunction. [Right] Chromosomes in late metaphase, the centromere clearly binds the separated sister chromatids.

### III. RESULTS

The imaged early metaphasic human chromosomes revealed the typical G-banding pattern as seen using AFM [Cummings, 2003]. G-positive bands were clearly seen at micrometric scale [fig 1 & 2]. The width for chromosome one was 1.40 ± 0.1 μm, which is in agreement with other AFM measurements for early metaphasic chromosomes [Fritzsche *et al*;1997].

The banding pattern accomplished with the AFM height images matched the light microscopically banding and the *Ensembl´s* ideogram [*Ensembl; 2008*] for chromosome one [Fig. 1]. Height graphics were chosen because they express the actual height and the compressibility of the sample being tested, providing information about the configuration and packing of the chromosome [Thalhammer *et al.* 2000; García & Pérez, 2002].

The longitudinal height graphics were quantitatively different depending on whether they were measured on one of the nascent chromatids (longitudinal zones 1 & 2) or in the middle of the chromosome (longitudinal zone 2). A pattern of ridges in the middle of the early chromosomes was observed in the longitudinal measurements. The transversal height graphics displayed this particular ridge always in the central region of the chromosome [fig. 1 & 3].

The G-positive bands were significantly higher than the negative ones ($p < 0.01$, T-test) thus, heterochromatic regions were thicker and more symmetric than the G-negative euchromatic regions [fig.3]. Interestingly, the mean values for heterochromatin heights (*Average 110 nm, Standard deviation 20 nm*) were around two folds larger than those for euchromatin (Average *54 nm, Standard Deviation 19 nm*). Under our experimental conditions the height ratio between heterochromatin and euchromatin (***H/E*** height ratio) was around 2:1 for most of chromosomes. We used this value as a rough indicator to asses the level of trypsin digestion, and only chromosomes with a 2±10 **H/E** height ratio were included in calculations, this range excluded around 10% of all the cases.

### IV. DISCUSSION

The longitudinal height measurements for the chromosomes displayed compact and loose regions. The white ridges on the image can be distinguished as peaks in the height graphic, while relaxed regions are showed in dark colors [fig. 2]. A comparison of the distribution of ridges and the grooves on the chromosome one, either from the AFM image or from an *Ensembl* ideogram, has a coincident euchromatin and heterochromatin patterns [fig. 1] according to the distribution of known genes.

The chromosomes were examined for early metaphase, in which the sister chromatids are not completely disjointed [fig. 2] although they can be realized as having almost their own conformations [Tamayo & Miles, 2000; Sumner, 1991]. The longitudinal height measurements along each of the sister chromatids and in the middle of the chromosome, showed that the higher points were predominantly located in their central longitudinal axis (fig 3). This behavior was previously realized while examining other samples for AFM in our laboratory. It was then confirmed in metaphasic spreads from different volunteers using chromosomes with an *H/E* height ratio close to 2. This ballpark parameter shows that euchromatin and heterochromatin were attacked in a proportional extent in the analyzed chromosomes, thus allowing a comparison of the chromosomes` structure after staining treatment.

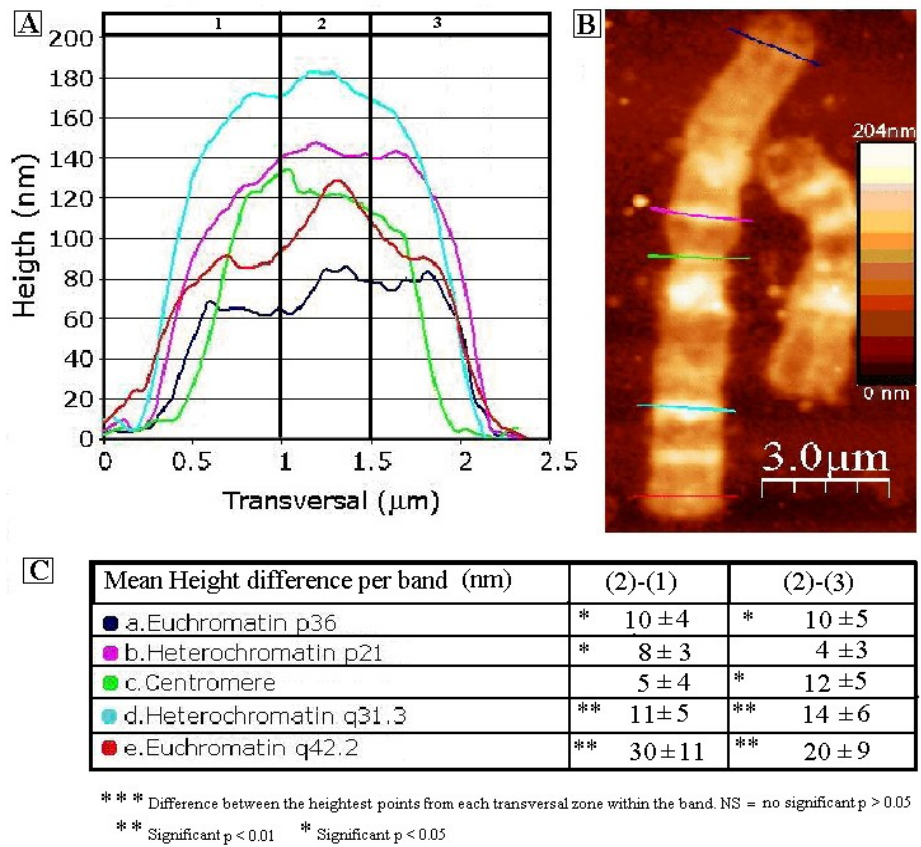

**Figure 3**. A) Transversal height sections for euchromatic and heterochromatic bands in chromosome one B) Topographic image of a representative G-banded metaphasic human chromosome one C) Averages of height differences tested against a null difference with equal sample variance.

The G-banding treatment brought a symmetrical transversal pattern across the chromosomes, which reaches its highest points at the center [fig. 3]. The fact that this chromosomal configuration was found in metaphasic spreads from different volunteers is evidence for its generalized presence at early metaphase.

This symmetrical transversal pattern in early metaphasic chromosomes could not be expected, since the trypsin treatment eats away the chromosome from the outside and one feasible option would be that each sister chromatid has its own ridge at its own center. Another possibility would be that the chromosome doesn't have any transversal pattern when is explored along the same band, i.e. having a rather flat height level across the sister chromatids. Neither of the mentioned options seems to correspond with our observations.

The chromosome's height pattern could be influenced by external factors like the deposition of stain, irregularities on the slide surface, remaining biological layers and sample debris from the preparation process [Harrison *et al*, 1982; Stark *et al*, 1998; Tamayo *et al*. 1999; Tamayo & Miles, 2000]. To discard the above cited possibilities we checked for irregularities in the glass support that might have biased the height measurements, but it was fairly flat in comparison to the chromosomes` heights [Fig. 3]. We also considered a remaining protein and RNA layer over the chromosome, which is about is around 30 nm height [Tamayo *et al.* 1999; Harrison *et al.* 1982], but since we measured height differences even of 100 nm along the chromosome our observations can not be explained by artifact layers. Furthermore, a biological debris layer from the preparation process could also be deposited on the glass substrate along with some Giemsa stain and produce slopes in the chromosome's disposition on the glass substrate; however in our preparations this layer had a typical value smaller than 20 nm. Therefore the differences in transversal and longitudinal height measurements, observed under our experimental conditions, were generated by intrinsic configurations of the chromosome after the G-banding treatment.

The enzymatic trypsin digestion of the chromosomes has been argued to produce the G-bands due to the depletion of non histones proteins, or just the degradation of the proteins located at the most loose parts of the chromosome [Kenneth *et al.* 1986; Hoshi & Ushiki, 2001; Rasch

*et al.* 1993]. In either case a differential arrangement of not only proteins, but also DNA is conceded. According to this, the transversal ridge and the symmetry pattern observed in our samples correspond to a specific arrangement of Proteins and DNA at a specific point of the metaphase. Despite their common central ridge, the transversal height graphics displayed a more symmetrical pattern in heterochromatic zones, as for the bands p21 and q31 [fig. 3.b, 3.d]. On the other hand, the euchromatic bands, like p36 and q42, tended to appear as having several peaks across the chromatids [fig. 3.a, 3.e]. Thus the heterochromatic bands accounted for the highest and more symmetric regions.

The observation of these ridges in the middle of the Chromosome and the pattern of transversal symmetry, can show an interaction between the sister chromatids prior to their separation. It has been observed that "interchromatidal interactions" take place in a noticeable way among heterochromatic bands [Grigoryev, 2004]. The presence of bridging fibbers, for instance, has been often observed in the heterochromatic bands [Wu *et al.* 2006; Ushiki *et al.* 2002]. There is evidence for specific DNA-protein arrangements that holds the sister chromatids during metaphase avoiding improper orientation or disjunction; one example of such arrangement is the phylogenetically conserved multiprotein complex called cohesin in humans, and Rad21 or Rec8 in plants and mice respectively [Zhang *et al* 2004] . There are also other proteic components that could be potentially associated with interchromatidal interactions in heterochromatin, and that could remain present after G-treatment due to a more compact arrangement. They might be proteins with similar functions to MeCP2 [methyl-CpG-binding protein 2], HP1 [heterochromatin protein 1], MENT [myeloid and erythroid nuclear termination stage-specific protein] among others; all of them interact strongly with DNA and/or histones [and their respective methylated, phosphorylated or acetylated variations] in order to organize spatial packing patterns in heterochromatin [Georgel *et al.* 2003; Nasmyth, 2003; Zhang *et al.* 2004; Kudo, 1998]. Thus, the relative great height and symmetry of the heterochromatic regions observed in this study could be thought as an indirect evidence for a differential arrangement of proteins and DNA, which can originate the posterior formation of structures like the bridging fibers observed in G-banded chromosomes. These results also support the role of heterochromatic regions in chromosome's stability during the process triggering the formation of sister chromatids from replicated chromatin.

Secondary peaks in both chromatids accompanied the central ridge in the Euchromatic bands [fig.3]. It can be theorized that these peaks display the way the chromatin is being distributed between the sister chromatids, and thus this direct observation indicates that the disjunction happens first in the euchromatic bands, which lose symmetry first. This confirms other results in the literature, which stress that disjunction is triggered first in euchromatic bands [Grigoryev, 2004].

According to our observations, a transitional chromosome arrangement leading towards the correct separation of the sister chromatids would be likely. This process would imply the primary condensation of chromatin, as seen in prophase; a secondary transitional state, mediated by a symmetrical distribution of DNA with associated proteins and the final disjunction of the chromatids, delayed specially in the heterochromatic bands which contribute to the chromosome's stability during this process.

The normal preparations for chromosome structure studies come from "mature" metaphasic chromosomes with their sister chromatids already split. We would like to suggest with our work that other stages of the chromosomal assemblage, like early metaphasic conformations, should be structurally distinguished from mature chromosomes in AFM studies, since these stages could determine the final structure of the chromosome. After all, if we want to figure out the development and structure of chromosomes we need to understand a continuous process in time, not an quasi-static accumulation of molecules with very discrete transitions.

## V. CONCLUSION

In this paper, we described chromosomal features which suggest that a transitional state of chromosomes prior to the splitting of the sister chromatids was captured by Atomic Force Microscopy (AFM), a method that has the advantage of obtaining the quantitative height and compressibility information of adsorbed chromosomes. We concluded that the AFM characterization of early metaphasic chromosomal features, such as the symmetry pattern seen in human chromosome one, can give hints on the processes undergoing the separation of sister chromatids.

### Acknowledgements

We thank Ortiz E, Obando C, Venegas P, from the laboratorio de citogenética del HNN for kindly provide the samples, also Madrigal-Paniagua C, Vidaurre-Baraona D, Lopez-Fernandez M, Bolaños N, for their support.